\documentclass[prb,aps,11pt,reprint,showkeys,superscriptaddress]{revtex4-2}

\usepackage{amsmath}
\usepackage{amssymb}
\usepackage{graphicx}
\usepackage{pgfplots}
\usepackage[caption=false]{subfig}

\usepackage[normalem]{ulem} 
\usepackage{color} 

\newcommand{\tocite}[1]{\textcolor{green}{\textit{\textbf{ cite}}}}

\pgfplotsset{compat=1.15}

\begin{document}
	
	
	\title{Microscopic quantum point contact formation as the electromigration mechanism in granular superconductor nanowires}
	\author{T. C. Bartolo}
	\email{tommy.bartolo@gmail.com}
	\affiliation{Chemical and Quantum Physics, School of Science, RMIT University, Melbourne, Australia}
	\author{J. S. Smith}
	\affiliation{Chemical and Quantum Physics, School of Science, RMIT University, Melbourne, Australia}
	\author{Yannick Sch\"on}
	\affiliation{Physikalisches Institut, Karlsruher Institut f\"ur Technologie, Karlsruhe, Germany}
	\author{Jan Nicolas Voss}
	\affiliation{Physikalisches Institut, Karlsruher Institut f\"ur Technologie, Karlsruhe, Germany}
	\author{M. J. Cyster}
	\affiliation{Chemical and Quantum Physics, School of Science, RMIT University, Melbourne, Australia}
	\author{A. V. Ustinov}
	\affiliation{Physikalisches Institut, Karlsruher Institut f\"ur Technologie, Karlsruhe, Germany}
	\affiliation{Institute for Quantum Materials and Technologies, Karlsruher Institut f\"ur Technologie,  Karlsruhe, Germany}
	\author{H. Rotzinger}
	\affiliation{Physikalisches Institut, Karlsruher Institut f\"ur Technologie, Karlsruhe, Germany}
	\affiliation{Institute for Quantum Materials and Technologies, Karlsruher Institut f\"ur Technologie,  Karlsruhe, Germany}
	\author{J. H.  Cole}
	\email{jared.cole@rmit.edu.au}
	\affiliation{Chemical and Quantum Physics, School of Science, RMIT University, Melbourne, Australia}
	
	\begin{abstract}
		
		Granular aluminium is a high kinetic inductance thin film superconductor which, when formed into nanowires can undergo an intrinsic electromigration process. 
		We use a combination of experimental and computational approaches to investigate the role of grain morphology and distribution in granular aluminium thin films, when formed into nanowire constrictions. 
		Treating the granular aluminium film as a network of randomly distributed resistors with parameters motivated by the film microstructure allows us to model the electrical characteristics of the nanowires. This model provides estimates of the dependence of sheet resistance on grain size and distribution, and the resulting device to device variation for superconducting nanowires.
		By fabricating a series of different length nanowires, we study the electromigration process as a function of applied current, and then compare directly to the results of our computational model. In doing so we show that the electromigration is driven by the formation of quantum point contacts between metallic aluminium grains. 
		
	\end{abstract}
	
	\maketitle
	
	\graphicspath{"./images/"}
	
	\newpage
	
	\section{Introduction} \label{section:intro}
	
	Nanowires composed of granular aluminium have recently been the subject of active research due to their high kinetic inductance.
	This allows for their integration in applications such as microwave kinetic inductance detectors\cite{Santavicca2016,Day2003,Mauskopf2018,Connor2019,Guo2017}, sensing or switching devices, constructing SQUIDs\cite{Deutscher1975}, the scaling of superconducting resonators\cite{Santavicca2016,Muller2018,Maleeva2018} and the formation of superinductors for use in qubits\cite{Wang2019,Grunhaupt2019,Kamenov2020}.
	Their high kinetic inductance and relatively low geometric inductance also makes them an ideal candidate for the study of quantum phase slips (QPS)\cite{Zgirski2008,Astafiev2012,Lehtinen2012a,Lehtinen2012b,Singh2013,Mooij2006}.
	In particular, new fabrication techniques for producing uniform, reproducible granular aluminium materials which have kinetic inductance values greater than 2 nH per square have been demonstrated\cite{Rotzinger2017,Voss2021}. Even more interesting, especially from a technological point of view, is that granular aluminium nanowires have been shown to undergo significant and irreversible resistance changes due to intrinsic electromigration under applied currents\cite{Voss2021}.
	
	Although the physics of granular superconductors has been studied for many years\cite{Parmenter1967,Deutscher1974,Abeles1975,Chui1981}, relatively little is known quantitatively about the role of the material's morphology and how this depends on growth conditions. 
    This is particularly true for granular aluminium which has a qualitatively different morphology\cite{Pettit} than other well studied 2D disordered superconductors such as MoGe\cite{Graybeal1987} or InO\textsubscript{x}\cite{Astafiev2012}.
    The surprisingly uniform microstructure of granular aluminium films gives us a unique opportunity to develop a meaningful model to explore the relationship between morphology and electrical response.
    This in turn provides a greater understanding of how the fabrication of granular aluminium nanowires impacts their performance as nanowire and QPS devices\cite{Vissers2015,Rotzinger2017,Astafiev2012,Kafanov2013}. 
	
	\begin{figure}[t]
		\centering
		\includegraphics[width=\linewidth]{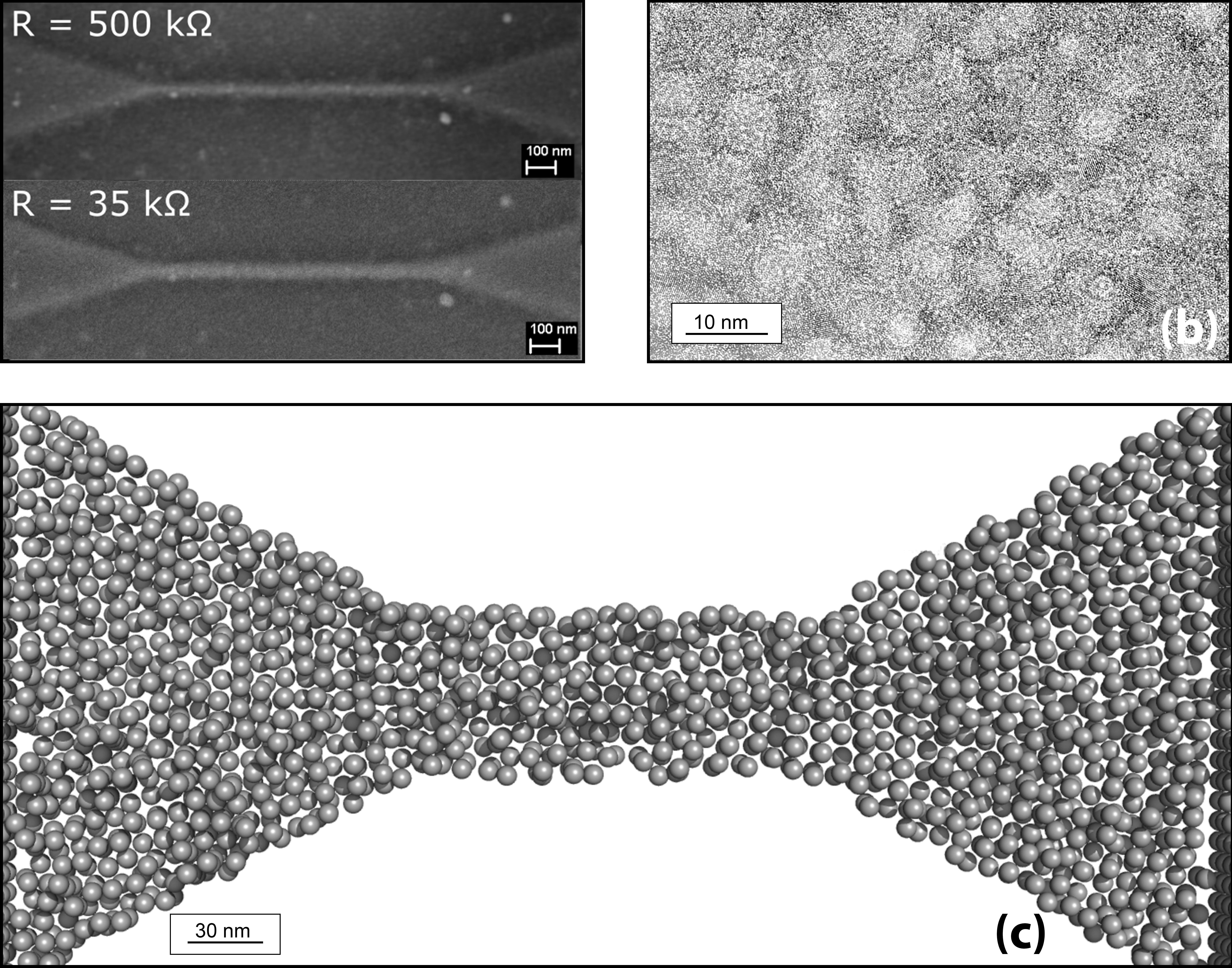}
		\caption{
			(a) Scanning electron microscopy image of a granular aluminium nanowire, before and after applied current. 
			(b) Transmission electron microscopic image of a granular aluminium thin film~\cite{Rotzinger2017}. 
			(c) A computer generated model of a nanowire granular aluminium channel, where each sphere represents a region of crystalline aluminium metal embedded within an aluminium-oxide dielectric.
		}
		\label{fig:granAl}
	\end{figure}
	
	The nanowires studied here were fabricated from an aluminum-oxide thin film, which was deposited by a reactive sputter process on a sapphire substrate\cite{Rotzinger2017}. Subsequently the sample is patterned by a reactive ion etch process, using an electron beam defined resist mask. Details of the process can be found in Refs.~\onlinecite{Yannick2020,Voss2021}.
	An example of these nanowires can be seen in Fig.~\ref{fig:granAl}a, which shows an scanning electron microscopy image of a granular aluminium nanowire before and after a current is applied to induce a change in the total resistance. There is no obvious change in the shape of the nanowire in this microscopy image, suggesting that the reduction in resistance is driven by modification of the internal microstructure.
	
	Fig.~\ref{fig:granAl} (b) shows a transmission electron microscope image of a granular aluminium film which is comprised of grains (regions of circular transverse cross-section in Fig.~\ref{fig:granAl}b) of crystalline aluminium metal, surrounded by a insulating aluminium-oxide dielectric.
	By visual inspection of the electron micrograph we estimate these grains are approximately 5--10 nm in diameter\cite{Rotzinger2017}. 
	The grains form a disordered network and conduction is assumed to proceed via tunnelling from grain to grain through the oxide dielectric\cite{Rotzinger2017,Abeles1975}. Further, the data from transmission electron microscopy suggests that as well as having a circular cross-section, the grains are fairly mono-disperse with little variation in diameter.
	
	These observations of the film microstructure motivates us to develop a computational model for conduction in granular aluminium nanowires, an example of which is shown in Fig.~\ref{fig:granAl}c.
	We model the connections between the metallic grains as tunnel junctions, i.e.\ metal-insulator-metal junctions.
	This allows for the network of grains to be approximated by a network of resistors, from which a normal resistance for the nanowire can be obtained.
	Having the means of computing a normal resistance for the nanowires also allows for the further calculation of their kinetic inductance and sheet resistance, properties which are key to their function in modern devices.
	That the normal resistance of the films is found to be similar at both room temperature and cryogenic temperatures suggests that such a semi-classical model is justified.
	
	In section~\ref{section:classicalmodel} we explain how granular aluminium thin films can be modelled as a network of randomly distributed resistors. The properties of the film depend on the distribution of resistors in the network, whose parameters are then chosen to reflect the microstructure of the experimental films.
	Further, we extract the total resistance for channels constructed from our model which we then discuss in relation to material properties in section \ref{section:contact_resistance/sheet_resistance}. 
	In sections \ref{section:radius_shellthickness_shellvariance} and \ref{section:Nanowire_variance} we discuss how the electronic properties in turn influence the performance of devices comprised of this material. Finally in section~\ref{section:wire_conditioning} we develop a minimum model for the electromigration process and we compare this to detailed experimental measurements of nanowires of various lengths.
	
	\section{Network model of conduction} \label{section:classicalmodel}
	
	We develop a resistor network model using a randomly distributed growth algorithm seeded with a single point, which can generate a nanowire of arbitrary size with quasi-randomly distributed grains. 
	Motivated by the transmission electron microscopy data (Fig.~\ref{fig:granAl}b) we treat each grain as a volume of superconducting aluminium while the surrounding amorphous alumina is modelled as the barrier between grains, and assume that the grains are relatively mono-disperse with little to no agglomeration. 
	
	Differences in grain size can be introduced as an additional disorder parameter, however for the purposes of this work we only consider a mono-disperse set of grains, each of which is modelled as a sphere.
	Similarly, variation in film thickness can introduce localised variation in sheet resistance. However, experimentally the film thickness is found to be extremely uniform. Therefore unless otherwise stated each channel considered here has a film thickness of 200 \AA, in line with those fabricated in Ref.~\onlinecite{Voss2021}.
	
	To develop our network model, we generate a three-dimensional structure of grains with a known mean and variance in the grain separation.	The problem of efficient random sphere placement is in general an unsolved problem in computer science\cite{Williams2003}. However, several heuristic algorithms exist\cite{Hifi2009}.
	
	To construct our network of grains, we start with an origin (typically the centre of the nanowire). 
	We then randomly assign a separation distance to the next grain from a predefined Gaussian distribution with mean $D_{\rm{mean}}$ and standard deviation $D_{\rm{std}}$, where the angle around the origin grain is assigned from a similarly defined distribution (with mean and standard deviation set to $\pi/4$). 
	The third grain is then placed again some separation distance from the origin grain, but now randomly assigning an angle such that it does not overlap with the second grain. 
	This process is repeated until no more grains can be placed around the first (origin) grain. 
	The entire process is then repeated with each of the newly placed grains as the origin, again placing grains such that the separation distance reflects the Gaussian distribution but no grains overlap.
	The process is continued, while respecting the condition that no grains are placed outside the boundaries which define the nanowire. 
	Once the entire device region has been patterned, the final product is a network of grains in a two-dimensional nanowire geometry. 
	The process is then repeated for additional layers on top of the first, such that the vertical separation is also approximately described by a Gaussian with appropriate mean and standard deviation.
	\begin{figure}[b]
		\centering
		\includegraphics[width=\linewidth]{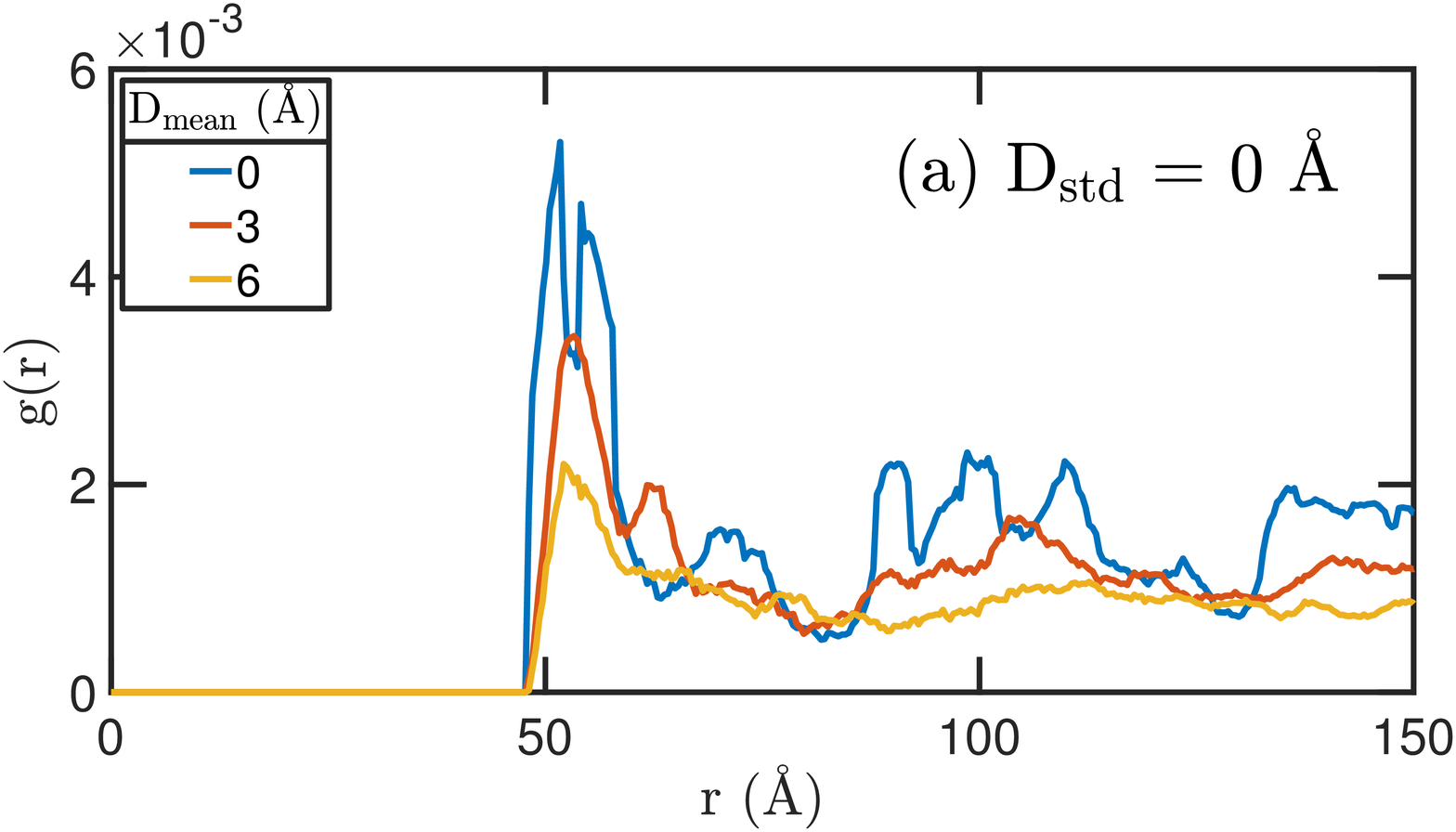}
			~
			\centering
			\includegraphics[width=\linewidth]{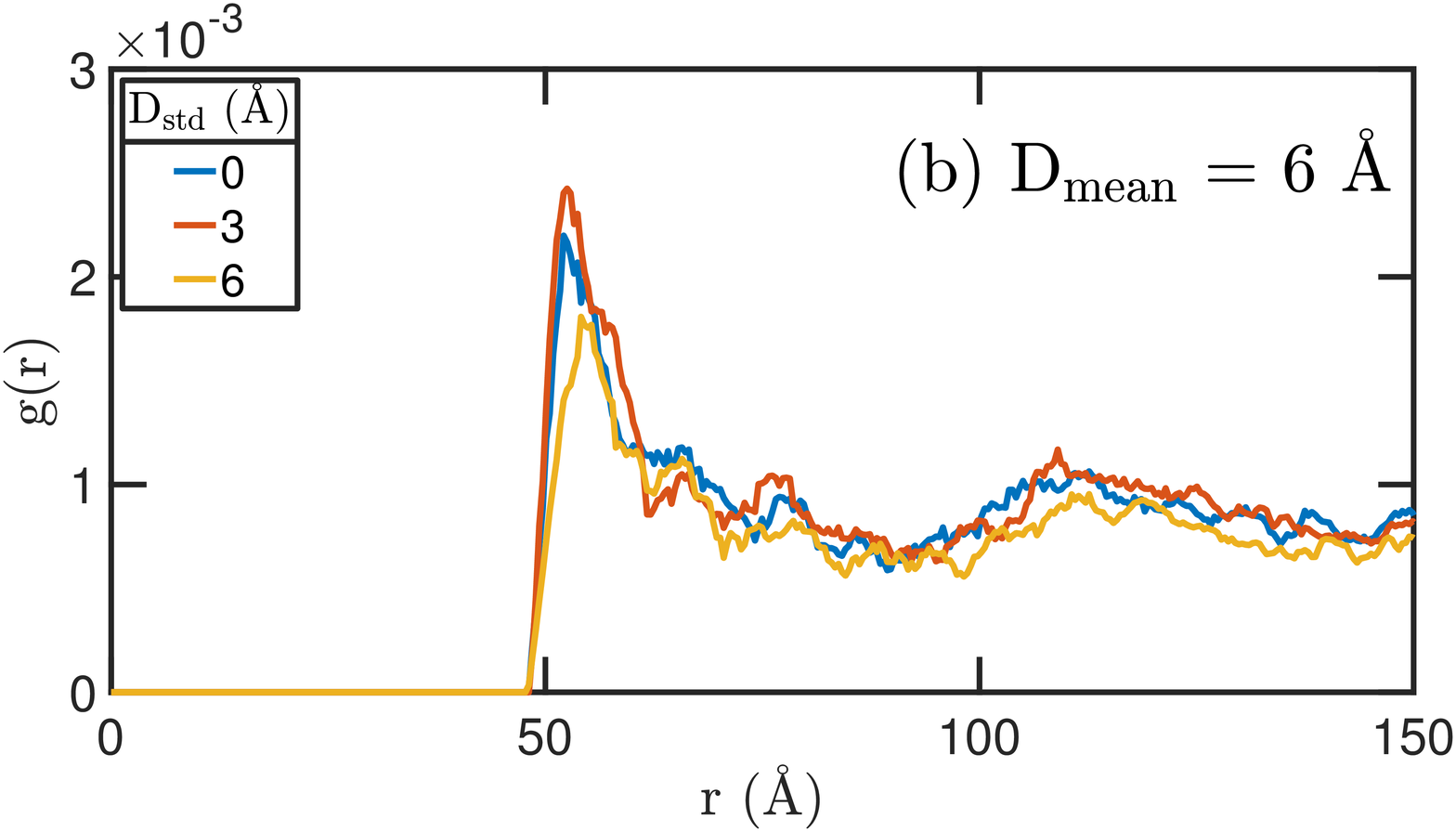}
			\caption{
				Radial distribution of grains in a simulated granular thin film as a function of mean separation. 
				(a) Shows the distribution for a standard deviation of 0~\AA~ for several different mean separations. 
				(b) Shows the distribution for a mean separation of 6~\AA~ for several different standard deviations.
			}
		\label{fig:disorder_wires}
	\end{figure}

	To confirm our algorithm produces a network that qualitatively reflects the input statistics, Fig.~\ref{fig:disorder_wires} shows the $g(r)$ computed for several values of $D_{\rm{mean}}$ with a standard deviation of 0~\AA (Fig.~\ref{fig:disorder_wires}(a)) and 6~\AA~(Fig.~\ref{fig:disorder_wires}b).
	For the case where both $D_{\rm{mean}}$ and $D_{\rm{std}}$ are zero, we find that the algorithm has inadvertently patterned the network in a hexagonal close packing structure. 
	However, as the mean separation and standard deviation are increased, we regain an amorphous structure with appropriate first nearest neighbour distance and peak width.
	In all that follows we assume a grain radius of 25~\AA, which is in keeping with previous estimates\cite{Abeles1977,Abeles1967,Cohen1968}.
	An example of a network of coupled grains patterned in a nanowire profile, with a typical distribution in separations, can be seen in Fig.~\ref{fig:granAl}c.
	
	The couplings between pairs of grains in a disordered superconductor are often treated as Josephson junctions~\cite{John1986,Bradley1988}.
	These Josephson junctions form links between superconducting regions, with an insulating material between them forming a tunnelling barrier\cite{Tinkham2004,Josephson1962}, i.e.\ the well-studied S-I-S junction.
	Each junction formed in this way has a characteristic critical current which can be used to determine the corresponding normal state resistance for the links via the Ambegaokar-Baratoff (AB) equation\cite{Ambegaokar1963,Ambegaokar1963errata}. 
	This relationship between the superconducting properties of the junctions between grains and the normal state resistance allows us to develop an effective resistor network model. 
	
	The sheet resistance of such a model can be determined by using the a graph Laplacian which is discussed further at the end of this section. Experimentally, the sheet resistance as measured at room temperature is a good predictor of the sheet resistance in cryogenic conditions. Similarly the intrinsic electromigration process when performed at room temperature was found to effectively control the resistance and the resulting low temperature response. These observations are the motivation for developing a effective semi-classical model for the resistor network, describing the sheet resistance and resulting nanowire resistance at room temperature.
	
	\begin{figure}
		\centering
		\includegraphics[width=\linewidth]{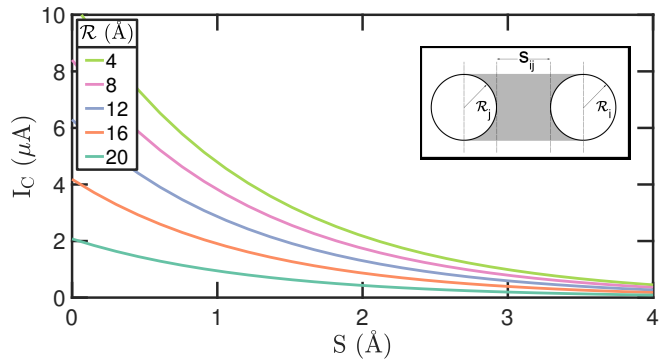}
		\caption{
			Critical current as a function of separation between two grains of crystalline aluminium metal separated by amorphous aluminium oxide.
		}
		\label{fig:EJvsSeparation}
	\end{figure}
	
	To estimate typical material parameters, we take advantage of the formalism laid out by Tinkham\cite{Tinkham2004}, for the critical current $I_c$ of a S-I-S type Josephson junction,
	\begin{align}
		I_c = \frac{2e\hbar\sigma\psi_\infty^{2}}{m^*}\exp\left(\frac{-S_{ij}}{\delta}\right)
	\end{align}
	where $m^*$ is the effective mass of the charge carriers. We use $m^* = m_e$ with $m_e$ being the mass of a free electron. 
	$S_{ij}$ is the separation between the grains which form the junction, $\delta$ is a characteristic constant of the potential barrier between grains, $\sigma$ is the interaction area between the grains and $\psi_\infty$ is the Ginburg-Landau complex order parameter which is used as a measure of the density of superconducting electrons in the bulk limit.
	
	We then follow Chudinov\cite{chudinov2002} in deriving an expression for the critical current. However in contrast with their work, we integrate over the full hemispheric surface of the neighbouring spheres to compute the Josephson energy between two spheres,
	\begin{align}
		I_c =  \frac{2e\hbar n_s}{m_e}\int_{0}^{\mathcal{R}} \int_{0}^{2\pi}  \exp\left(\frac{-S_{ij}}{\delta}\right)d\theta dr \label{eq:Ic}
	\end{align}
	giving
	\begin{equation}
		I_c = \left(\frac{\pi e \hbar n_s}{m_e}\right)\mathrm{exp}\left(\frac{-S_{ij}}{\delta}\right)\left[\left(2\mathcal{R}-\delta\right) + \delta \mathrm{exp}\left(\frac{-2\mathcal{R}}{\delta}\right)\right] \label{eq:EJchudinov}
	\end{equation}
	where $\hbar$ is the reduced Planck's constant, $n_s = 2.5\times 10^{-5}$ is the condensate fraction of aluminium metal\cite{Salasnich2010} and $\mathcal{R}$ is the radius of the grain.
	The result of Eq.~\ref{eq:EJchudinov} is an exponential decay in the Josephson energy as a function of the separation between two grains as can be seen in Fig.~\ref{fig:EJvsSeparation}.
	
	For the purposes of linking critical current and normal resistance, we take the low temperature limit. The observed critical temperature varies depending on sheet resistance, but for the wires studied here is in the range  $T_c=1.6-2.0$~K. However it is important to note that the intrinsic electromigration process discussed in section~\ref{section:wire_conditioning} was found to be effective at both cryogenic and room temperatures~\cite{Voss2021}. This suggests that our semi-classical resistance model is able to describe the electromigration process even though it does not describe the movement of Cooper-pairs or magnetic flux quanta in the superconducting network below the critical temperature. 
	
	We define $\delta$ as the characteristic constant of the dielectric material which forms the barrier between the grains, in this case, aluminium oxide\cite{chudinov2002} which is calculated from the potential barrier and is given by $\delta = \hbar/\sqrt{2 m^* U_0}$.
	In this work we assume that potential barrier of a aluminium-oxide junction is ${U_{0} = 2.3}$ eV which lies within previous experimental estimates of approximately 1.8--3 eV depending on density and stoichiometry of the aluminium-oxide\cite{Sloof2002,Park2002,Cyster2020}.
	
	We then convert between the Josephson energy of a junction and its corresponding normal state resistance by way of the AB equation, \cite{Ambegaokar1963,*Ambegaokar1963errata,Abeles1977,chudinov2002}
	\begin{align}
		R_N = \frac{\pi\Delta_0}{2e I_c}
	\end{align}
	where $R_N$ is the normal state resistance of a junction, and $\Delta_0$ is the BCS superconducting gap at zero temperature.
	
	This model is equivalent to computing the conductance due to normal electrons tunnelling through the dielectric barrier between two metal hemispheres. 
	We simply express this in terms of $I_c$ to make a connection to existing models in the literature for conduction in granular superconducting films and the parameters of Al/AlOx/Al junctions.
	
	\subsection*{Computing network resistance via the graph Laplacian}
	
	To compute the electrical response of a disordered network of resistors, we follow Wu \cite{Wu2004} in constructing graph Laplacians and extracting resistance calculations from them. The Laplacian matrix consists of two components, a degree matrix which can be thought of as on-site energy terms, and an adjacency matrix which represents the connectivity of the grains. 
	These elements are constructed from the edge set $E$, which contains 2-tuples built from the vertices of the graph $V$, such that for each pair of grains $i$ and $j$ we have, $(i,j)\in E$.
	These edges represent the magnitude of the coupling between any two grains, which are represented by the vertices of the graph.
	The conversion between the indexed set of edges and a matrix is a completely natural one, where the indices of each element map to rows and columns in matrix form. 
	In this way we can build an adjacency matrix,
	
	\begin{align}
		A_{ij} &= (r_{ij})^{-1} \\
		\intertext{with}
		c_{ij} &= r_{ij}^{-1} = c_{ji}.
	\end{align}
	
	where $c_{ij}$ are the elements of the edge set and are equivalent to the conductance between the $i$\textsuperscript{\textit{th}} and $j$\textsuperscript{\textit{th}} grains in this formalism.
	The adjacency matrix is built such that none of the grains are capable of coupling to themselves and therefore the 2-tuple $(i,i)\notin E, \hspace{0.2cm} \forall i\in V$.
	
	The degree matrix is a matrix where each element corresponding to a certain grain contains the sum of the magnitudes of all coupling matrix elements associated with that grain.
	When using an adjacency matrix whose elements are either 0 or 1, the degree matrix will simply have the number of connections to each grain as its main diagonal elements. 
	The elements of the degree matrix are obtained by summing the weighting of each edge attached to each grain individually,
	
	\begin{equation}
		D_{i\alpha} =  \delta_{i\alpha}\sum_{j}^{N}A_{ij} \label{degree_equation}
	\end{equation}
	where $\delta_{i\alpha}$ is the kronecker delta.
	
	Considering our adjacency matrix to be the kinetic terms which couple our grains and the degree matrix to be an on-site energies, we may write the graph Laplacian as,
	
	\begin{equation}
		\mathcal{L} = D - A \label{LDA}
	\end{equation}
	
	such that we have
	
	\begin{center}
		\begin{math}
			\mathcal{L} =
			\begin{bmatrix}
				~c_{11} & -c_{12} & -c_{13} & \cdots & -c_{1n} \\
				-c_{21} & ~c_{22} & -c_{23} & \cdots & -c_{2n} \\
				-c_{31} & -c_{32} & ~c_{33} & \cdots & -c_{3n} \\
				\vdots &     \vdots  &  \vdots  &     \ddots     &  \vdots \\
				-c_{m1} & -c_{m2} & -c_{m3} & \cdots & ~c_{NN}
			\end{bmatrix}.
		\end{math}
	\end{center}
	
	Diagonalizing this Laplacian allows us to determine the resistance between any two points of the channel, which we label $\alpha$ and $\beta$,
	
	\begin{equation}
		R_{\alpha\beta} = \sum_{i=2}^{N} \frac{1}{\lambda_i}\vert \psi_{i\alpha} - \psi_{i\beta}\vert^2. \label{eq:GL_resistance}
	\end{equation}
	
	where $\lambda_i$ and $\psi_{ix}$ are the $i$\textsuperscript{\textit{th}} eigenvalue and $x$\textsuperscript{\textit{th}} element of the $i$\textsuperscript{\textit{th}} eigenvector of the Laplacian respectively. The symmetry of the Laplacian\cite{Shafarevich2013} ensures the first eigenvalue $\lambda_1 = 0$.
	
	\section{Contact and sheet resistance} \label{section:contact_resistance/sheet_resistance}
	
	In studying a finite channel there will be grain placement near the boundaries of the system which may perturb the resistance of the nanowire.
	One such region where edge effects become problematic is at the interface between the nanowire and the leads.
	To account for these boundary effects we broaden the nanowire at this interface, which increases the surface area between the nanowire and the contacts in our model.
	An example of this can be seen in Fig.~\ref{fig:granAl}c.
	The increased cross-sectional area at the interface which this broadening creates will reduce the effect of resistance perturbations due to the placement of grains at the boundary of our model.
	
\begin{figure}[b]
		\centering
		\includegraphics[width=\linewidth]{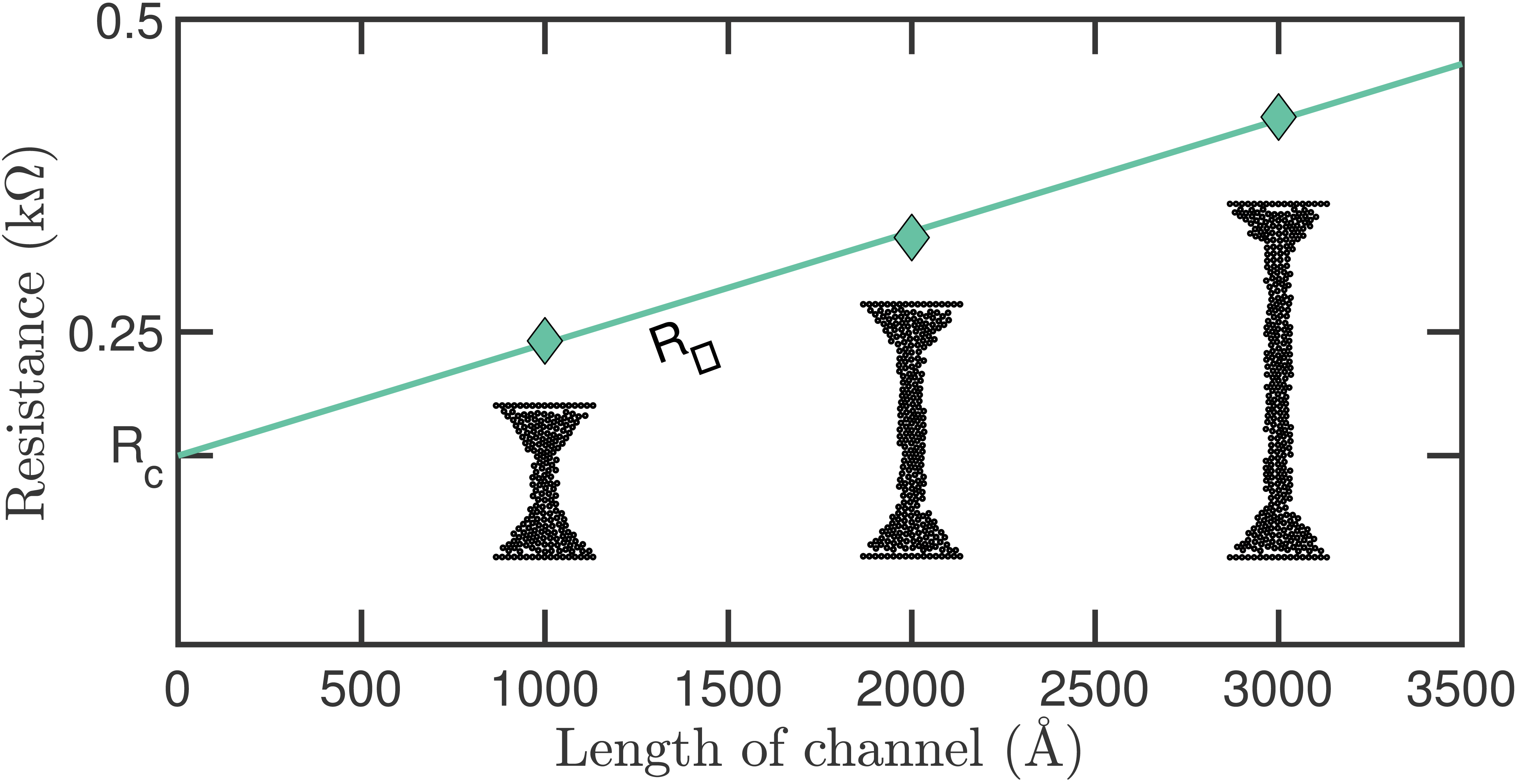}
		\caption{
			Example of the simulated normal resistance for several lengths of a channel. 
			The width of the channel is 1000 \AA. 
			Here the mean and standard deviation in the separation of the grains is 3 \AA.
		}
		\label{fig:extract_R_contact}
	\end{figure}

	The additional resistance contribution due to these `contact' regions can be subtracted from the total resistance, allowing an estimate of the resistance of the nanowire itself. 
	For all nanowires considered in this work, we set the size of this broadening region to be 1000~\AA~$\times$~1500~\AA~where the width of the device region is linearly reduced to the nanowire width at either end (see Fig.~\ref{fig:granAl}c).
	An example of how this contact resistance is calculated is shown in Fig.~\ref{fig:extract_R_contact} where the length of the channel is extrapolated down to zero and the $y$-intercept equates to the resistance due to the broadening of the channel-lead interface (the contact resistance $R_c$).

	\begin{figure}[t]
		\centering
		\includegraphics[width=\linewidth]{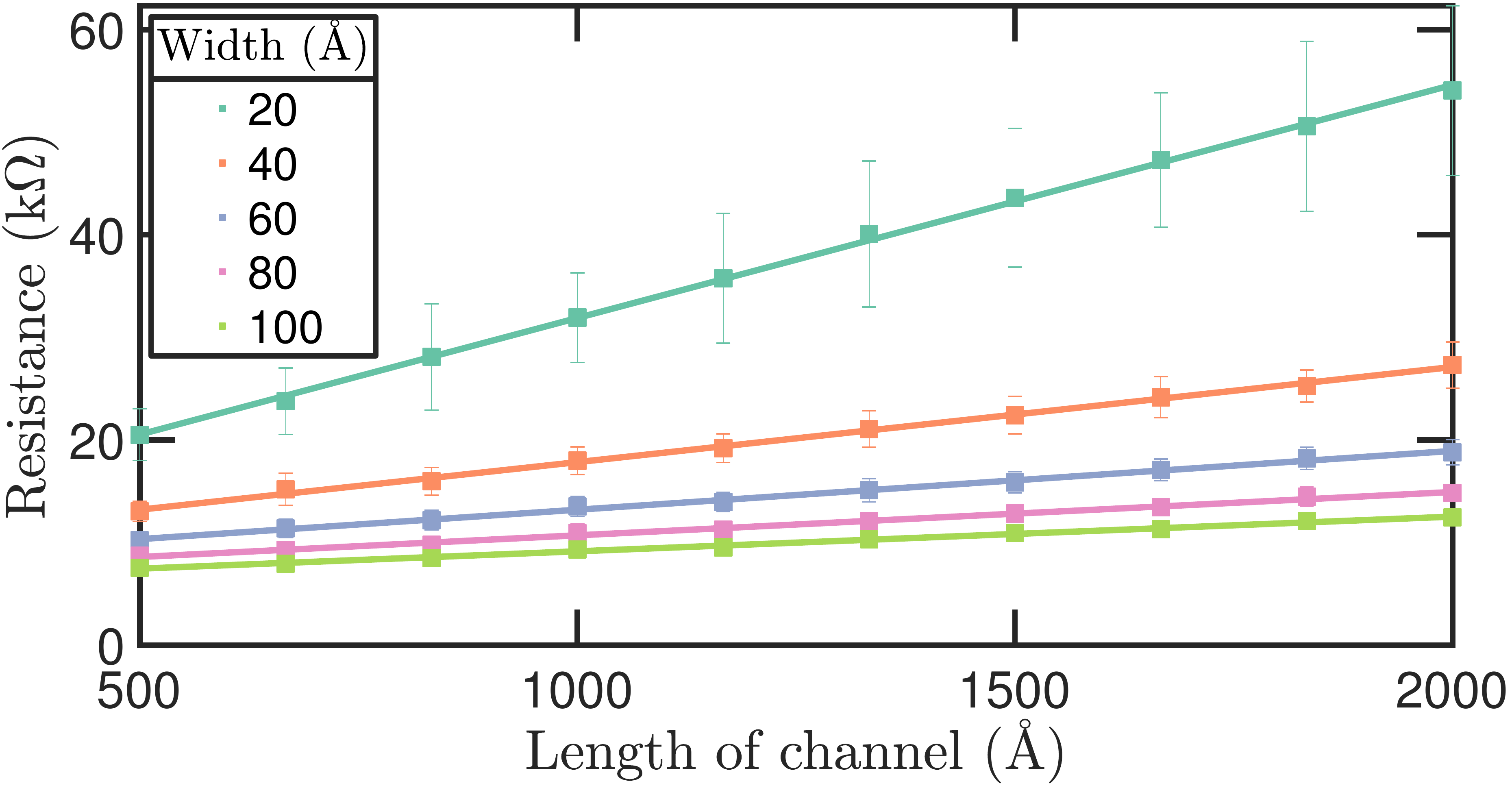}
		\caption{
			Simulated channel resistance versus the length as the width is varied for 100 realisations of a nanowire with the same input parameters.
			Here we use a mean separation and standard deviation of separation of 3 \AA.
			Error bars are taken from the standard deviation of the repeated iterations of the nanowire at each size, goodness-of-fit values for the trendline are all in excess of 0.99.
		}
		\label{fig:JM1}
	\end{figure}
	
	The sheet resistance of these channels is determined by calculation of the resistance as a function of channel length and width. 
	There is a strong linear dependence between the length and width of the channel and the resistance, as is shown in Fig.~\ref{fig:JM1}, which surveys many computationally generated channels with the same material properties. 
	This trend breaks down when the channel is narrow, where one or both physical dimensions are small enough such that fluctuations in the distribution of the channels results in large changes in their resistance.
	
	\section{Variation in material properties} \label{section:material_variance}
	\label{section:radius_shellthickness_shellvariance}

	Sheet resistance is a scale invariant property and therefore does not change with the physical dimension of a bulk material (i.e.\ length or width). 
	However it is affected by variations in the composition and structure of the material itself.
	Here we vary certain material properties of the nanowires, such as distribution of the grains, in order to see their effect on their electrical characteristics.
	
	If we consider a random network of metallic grains, there are three regimes depending on the variance in the separation of the grains. These are the low variation limit, where grains are distributed uniformly ($D_{\rm{std}}\ll D_{\rm{mean}}$), the sparse limit, where the variance in grain separation is so large it leads to open circuits ($D_{\rm{std}}\gg D_{\rm{mean}}$), and an intermediate regime ($D_{\rm{std}}\approx D_{\rm{mean}}$).
	
	For channels which are created in the low variance regime we see an oscillatory behaviour in the resistance. 
	This behaviour is due to low variance in the distribution causing a uniform spacing between the grains indicative of the hexagonal close packing seen in the $g(r)$ functions in Fig.~\ref{fig:disorder_wires}a. 
	This uniformity results in periodic fluctuations in the connectivity at the boundaries of the channel (and the resulting resistance), as the channel length is increased, which are non-physical.
	
	Applying our grain placement algorithm to channels with a large variance in the mean separation will create large spaces between each pair of grains. Since the coupling strength between each grain pair is exponentially suppressed with increasing separation, we see that complete circuits cannot be formed. 
	This causes a large increase in the resistance, tending toward open circuits, as the separation variance increases.
	It is therefore the intermediate regime that we focus on when comparing to the experimentally measured electrical response.
	
\begin{figure}[b]
		\centering
		\includegraphics[width=\linewidth]{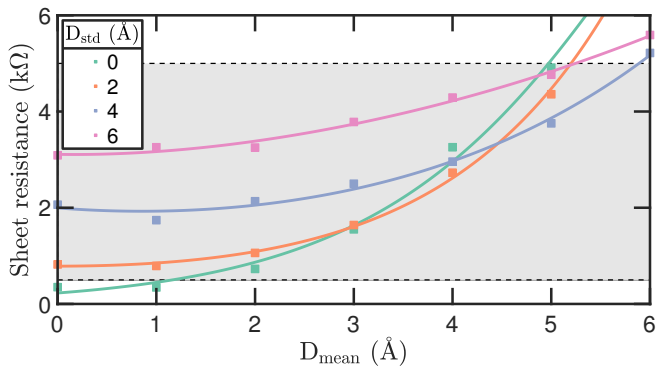}
		\caption{
			Simulated sheet resistance vs mean radius for a channel approaching the bulk limit with an exponential fit overlaid. 
			Here we use a channel width of 1000~\AA.
			The grey region represents the typical range of sheet resistance observed in experimental nanowires\cite{Rotzinger2017}. 
		}
		\label{fig:main_result}
	\end{figure}

	Fig.~\ref{fig:main_result} shows the sheet resistance for a bulk sample (where the length and width of the nanowire are much larger than the grain diameter) as a function of the mean and the standard deviation of the separation between grains.
	There is an exponential increase in sheet resistance as a function of the mean separation between grains.
	The exponential increase in sheet resistance follows from the exponential behaviour seen in the strength of the coupling between two grains as the separation is varied (Fig.~\ref{fig:EJvsSeparation}).
	Further, we also see an increase in the sheet resistance as the standard deviation in the separation between grains is increased.
	The range of observed sheet resistances from fabricated nanowires is highlighted in grey\cite{Rotzinger2017} in Fig.~\ref{fig:main_result}.
	From this highlighted area we can see the range of parameters for which our model matches experimentally expected behaviour.
	
	\section{Nanowire variance and its implication for QPS devices}\label{section:Nanowire_variance}
	
	A key problem with using high kinetic inductance nanowires for superconducting circuits is one of reliability and the resulting device yield\cite{Astafiev2012,Yannick2020,Belkin2015,Aref2012,Niepce2019,Arutyunov2012,Peltonen2013}. 
	The ideal device for observing coherent quantum phase slips has very narrow wires, made from a material with very high kinetic inductance. 
	However as these materials are disordered superconductors this is exactly the limit where device to device variation is large, making design and fabrication of working devices very difficult.
	This effect is pronounced in the situation where the nanowire becomes increasingly narrow and edge effects start to dominate over the contributions from the bulk of the channel.

	One of the advantages of the granular aluminium films considered here is that the high level of uniformity of the grains and the microstructure\cite{Pettit} suggests that the device to device variation could also be greatly reduced. 
	It is therefore of great interest to determine what type of variation is possible, even in an idealised model.
		
\begin{figure}[t]
		\centering
		\includegraphics[width=\linewidth]{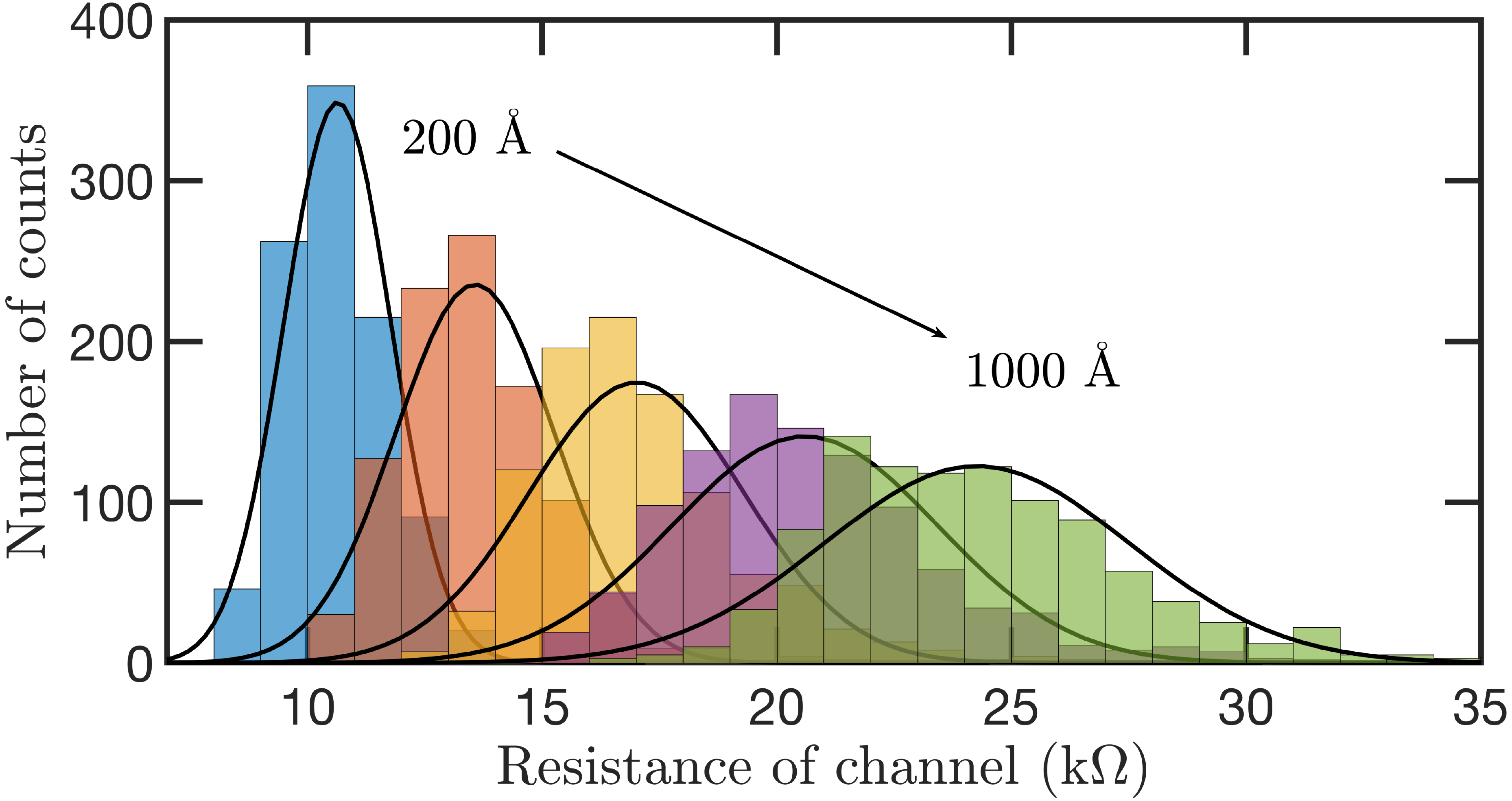}
		\caption{
			A histogram representation for the data in Fig.~\ref{fig:JM1} showing the (simulated) device to device variation. Each histogram is determined from 1000 independently generated nanowires, each with a width of 200~\AA~and with lengths varying from 200~\AA~to 1000~\AA~(in steps of 200~\AA).
		}
		\label{fig:Histogram_resistance}
	\end{figure}

	Fig.~\ref{fig:Histogram_resistance} shows many instances of a wire with the same input parameters and the variation in the distribution of resistances obtained.
	Here we have taken 1000 instances of nanowires with different lengths but otherwise possessing uniform physical characteristics (200 \AA~width as the length is varied from 200 \AA~to 1000 \AA, with $D_{\rm{mean}}=3$~\AA and $D_{\rm{std}}=3$~\AA). 
	From this we see that for a wire length of 1000~\AA, we can still expect a device to device variation in the resistance of $24$~k$\Omega$~$\pm$~$3.3$~k$\Omega$ (1 standard deviation). For a longer wire (1~$\mu$m) we obtain wire resistances of $486$~k$\Omega$~$\pm$~$66$~k$\Omega$ for the same nanowire width, but $D_{\rm{mean}}=3.5$~\AA~and $D_{\rm{std}}=3.5$~\AA.
	
	When modelling granular nanowires we observe examples in which current is unable to flow due to a large separation between the grains in any one path through the material. 
	This is especially prevalent in very narrow nanowires where the width is of order the variance in grain separation. In such a situation, there is a high chance of creating a nanowire which no longer conducts, or does so with a much higher resistance than intended -- which corresponds to the sparse regime discussed earlier.
	
	In an experimental setting, the possibility of unique clustering arrangements for the aluminium grains in granular aluminium nanowires must also be considered and will have a similar effect. 
	If grains display a tendency to cluster together as the aluminium content is increased in these films, this would drastically alter the electronic properties of the films, as electron transport is exponentially suppressed with grain separation.
	
	
	\section{Nanowire conditioning via electromigration} \label{section:wire_conditioning}
	
	\begin{figure}[tb]
		\centering
		\includegraphics[width=\linewidth]{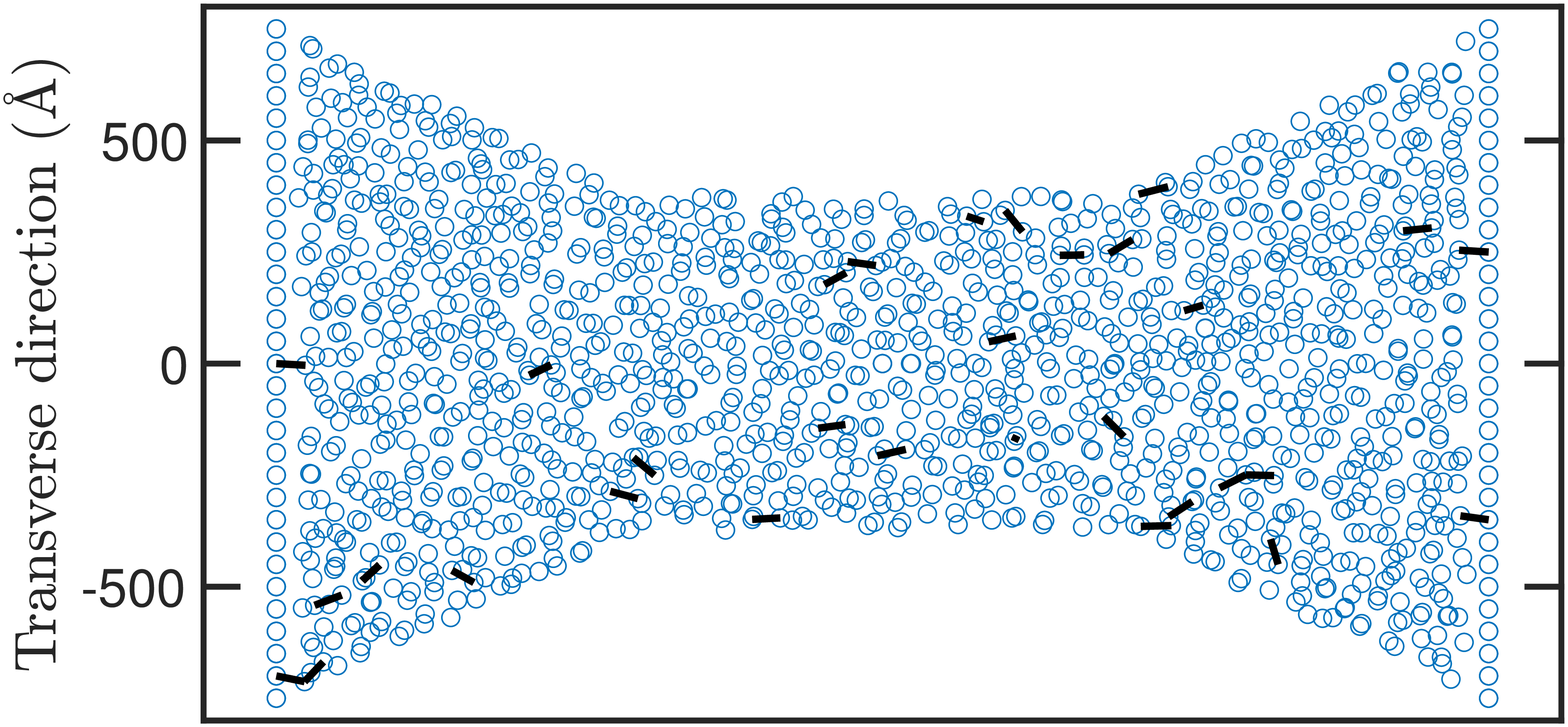}
		~
		\centering
		\includegraphics[width=\linewidth]{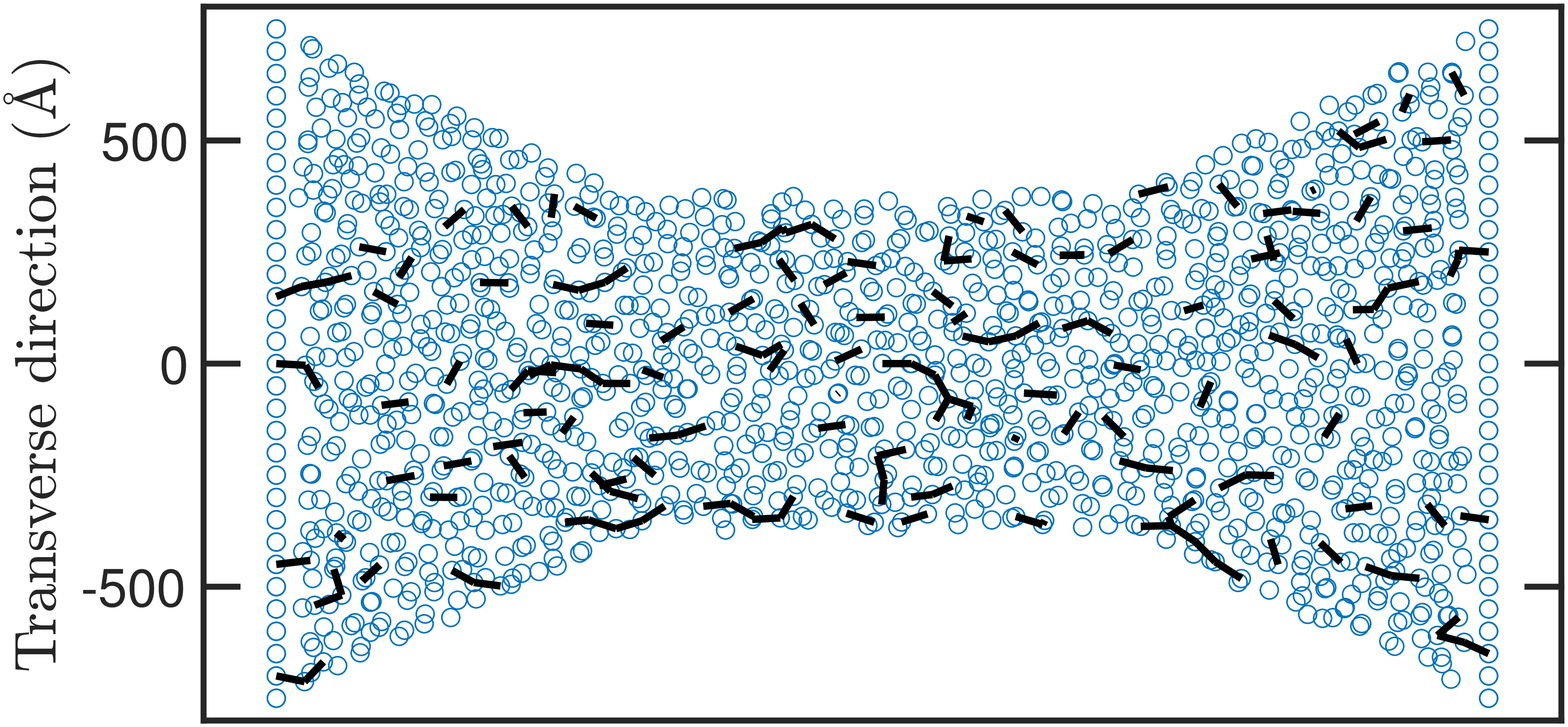}
		~
		\centering
		\includegraphics[width=\linewidth]{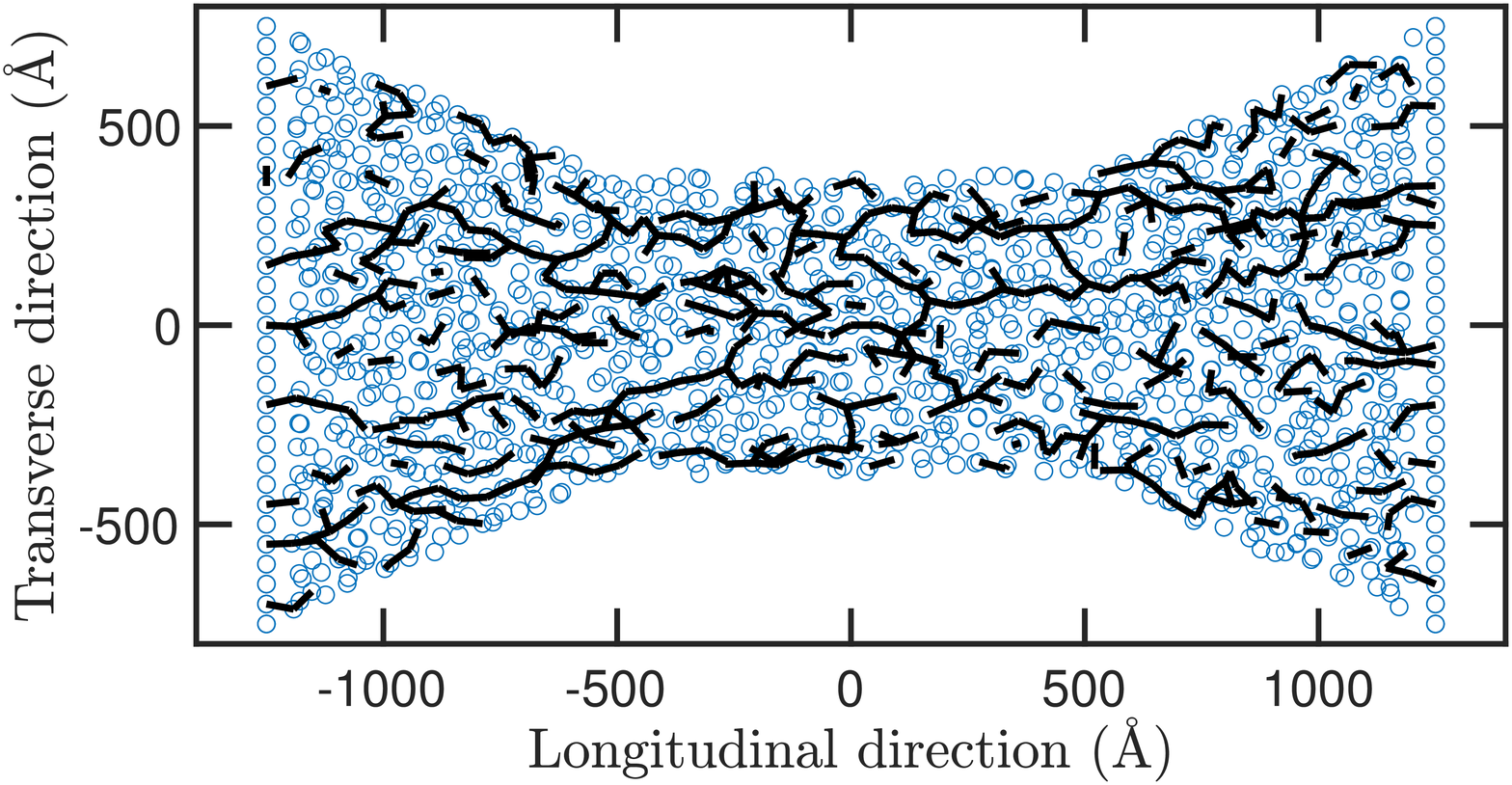}
		\caption{
			Representation of a 3D nanowire device with individual grains represented as blue spheres with different applied currents. 
			Here the wires have an applied current of a) 46 $\mu$A, b)  212 $\mu$A and c) 1000 $\mu$A, and new QPC links formed due to this current are represented as black connections between grains.
			This is a nanowire with dimensions of 1000~\AA, 750~\AA~and 100~\AA~in the longitudinal and (two) transverse directions respectively.
			Note that this is a top down view of the device which contains several layers of grains, and hence connections may be directed into or out of the page.
		}
		\label{fig:3_frame_melting}
	\end{figure}
	
	The movement of atoms within a material in response to an applied current, electromigration, has been well studied as a failure mechanism in electronic devices~\cite{Lloyd1997} and as a method of switching in memresistors~\cite{Strukov2008}. In this context, much is known about how the movement of defects affects the electrical response, including in aluminium devices~\cite{Lombardo2019, Guo2020}.	There have also been extensive studies of percolation in metallic nanoparticle clusters~\cite{MeiswesBroer2000,Milani1999}, where the electrical response can be modified by applying currents to induce preferred conduction pathways~\cite{Dunbar2006,Schmelzer2002, Sattar2013}. The granular aluminium films considered here appear to undergo similar percolation effects driven by electromigration.
	
	Recent work has demonstrated that the application of current pulses across a granular aluminium channel will induce an intrinsic electromigration (IEM) process\cite{Voss2021}. This technique works at room temperature or cryogenic temperatures and can be used to `condition' the nanowires to achieve precise control over their resistance as part of the fabrication process.
	It was proposed that this process merges neighbouring grains or grain clusters together, which in turn reduces the normal state resistance of that device\cite{Voss2021} -- similar to what is observed in metal nanoparticle clusters~\cite{MeiswesBroer2000,Milani1999}. However, it is not clear how to reconcile this picture with the very uniform and monodisperse microstructure seen in TEM. Do the grains grow? If so, where does the extra metal come from? Do they move? If so, why doesn't clustering result in open circuits in thin wires?
	Using our computational model of grain microstructure we find that the formation of localised conducting channels between the grains due to the movement of oxygen atoms is a more realistic model.	
	
	Based on detailed studies of Josephson junction breakdown, Tolpygo et al.\ found that the applied voltage across the barrier results in a movement of oxygen atoms in the oxide\cite{Tolpygo2008}. 
	This in turn allows the formation of conducting channels which form quantum point contacts (QPCs). 
	Here we assume a similar effect happens in the effective barriers formed between nanometre sized grains. We then show that this results in a response which correspond very closely to that seen in experimental measurements of granular aluminium nanowires.
	
	To model this process we follow Tolpygo et al.\ in simulating junction breakdown events as the creation of QPC links across junctions\cite{Tolpygo2008}. 
	The number of QPC links $N$ which are created depend on the separation and voltage $V_{ij}$ across any particular junction, given by
	
	\begin{align}
		N = B t \sinh(V_{ij}/V^{c}_{ij}) \label{eq:N_QPCs}
	\end{align}
	
	where
	
	\begin{align}
		V^{c}_{ij} = \frac{k_{\mathrm{B}}T S_{ij}}{qa}.
	\end{align}
	
	Here $k_{\mathrm{B}}$ is the Boltzmann constant, $T$ is the temperature, $S_{ij}$ is the barrier thickness between a grain pair, $q$ is the ion charge, $a$ is the activation distance, $B$ is a temperature dependent parameter and $t$ is the stress duration. Once the conductance of a link is increased due to the addition of $N$ QPC links, the voltage across that link will typically be reduced. The resulting effect is that Eq.~\ref{eq:N_QPCs} becomes a condition for forming a percolation link due to the formation of a metallic pathway between grains. The formation of these new links results in the overall reduction in the resistance of the nanowire.
	
	To determine the voltage drop across a junction ($V_{ij}$) which connects grains $i$ and $j$, we use a combination of Ohm's law, Kirchoff's current law (KCL) and the Moore-Penrose pseudoinverse\cite{Penrose1955}.
	
	We begin with a matrix formulation of Ohm's law,
	
	\begin{align}
		\hat{V} = \hat{I}\mathcal{L}^{-1}
	\end{align}
	
	where the Laplacian matrix $\mathcal{L}$ is obtained by taking the Moore-Penrose pseudoinverse of our resistance matrix.
	We can introduce a current across the device in the form of a current matrix which gives the net current flowing through each grain,
	
	\begin{align}
		\hat{I} = [I_0,0,\cdots,0,-I_0]
	\end{align}
	
	where $I_0$ is the magnitude of the current which is applied across the device. Here KCL ensures that for all grains except for the source or drain, the net current will be zero.
	We may now calculate the voltage drop across each junction as
	
	\begin{align}
		V_{ij} = I_0 \bigg((\mathcal{L}^{-1})_{i1} - (\mathcal{L}^{-1})_{iN} - (\mathcal{L}^{-1})_{j1} + (\mathcal{L}^{-1})_{jN}\bigg)
	\end{align}
	
	where the subscripts attached to the inverse Laplacian refer to the elements of the inverted matrix Laplacian.
	
	To determine which junction undergoes the IEM process, we use the small angle approximation and arrive at a simplified form of Eq.~\ref{eq:N_QPCs},
	
	\begin{align} \label{eq:parameter_A}
		N = \frac{Aq}{k_BT}  (V_{ij}/S_{ij}),
	\end{align}
	
	where $A = Bta$, which is used as a fitting parameter in what follows.
	Any junction with a value of $N>1$ will have $\lfloor N \rfloor $ (rounded down) QPCs created in parallel between the two grains which the junction connects.
	As the current is applied across the device QPC links are added until there exists no further junctions which fit the condition outlined by Eq.~\ref{eq:N_QPCs}.
	
	\begin{figure}[b]
		\centering
		\includegraphics[width=\linewidth]{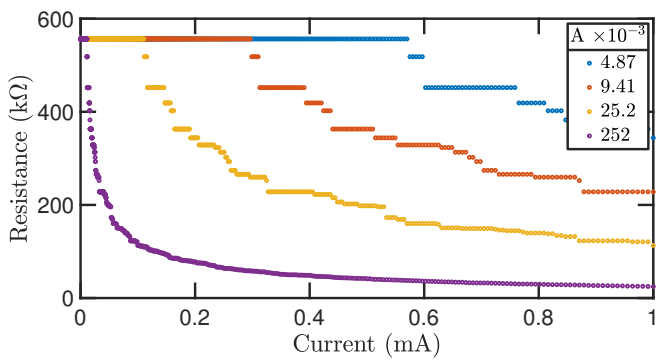}
		\caption{
			Simulated resistance of a nanowire as it is conditioned with an applied current, for several different values of the fitting parameter $A$. 
			Here we use a nanowire that is 1000 \AA~long, 750 \AA~wide and 100 \AA~thick.
		}
		\label{fig:wire_melt_A}
	\end{figure}
	
	The applied current can then be increased and the process is repeated for the now partially shorted device. 
	A visual representation of this process can be seen in Fig.~\ref{fig:3_frame_melting}, where a single nanowire device is shown for three different values of applied current.
	This shows a top-down look of the 3D device with the shorted junctions highlighted in black.
	As the applied current across the nanowire increases, shorted junctions begin to combine to form pathways through the device which significantly decreases the resistance of the nanowire, as is observed experimentally. Although the model introduced here is based on normal conduction, the formation of conduction pathways suggests that the superconducting transition seen in Ref.~\cite{Voss2021} appears once a continuous percolation pathway develops through the device, as can be seen in Fig.~\ref{fig:3_frame_melting}c. In this limit, the QPC formation between superconducting grains allow for coherent tunnelling of charges from one end of the wire to the other. This would result in an effective coherence length whose value is set by the superconducting network as a whole, rather than the properties of the individual grains.

	Determining how these wires will behave during the IEM process is critical to using this technique for device conditioning and obtaining nanowires with desired specifications.
	Fig.~\ref{fig:wire_melt_A} shows the the exponential decay in resistance due to increasing applied current, as the fitting parameter $A$ is modified. 
	As a current is applied to the device the creation of QPC links across junctions in the device leads to an overall lowering of the device resistance.
	Increasing the fitting parameter $A$ in Eq.~\ref{eq:parameter_A} has the effect of lowering the threshold current required to begin forming QPC links.
	This in turn induces an exponential shift in the threshold current required to condition the device and achieve a certain resistance. 
	
\begin{figure}[t!]
		\centering
		\includegraphics[width=\linewidth]{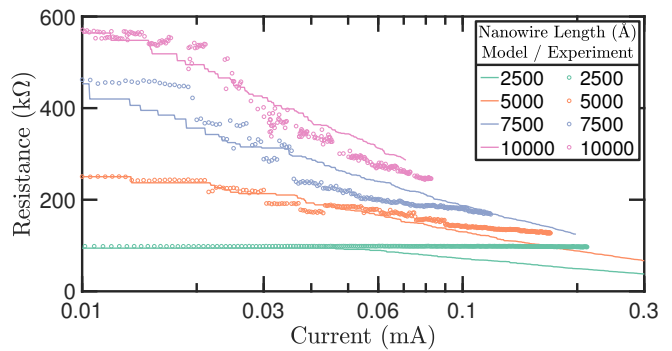}
		\caption{
			Resistance of experimentally (dots) and simulated (lines) nanowires after the application of a current to induce IEM for several different lengths of nanowire. The value of the IEM fitting parameter used for the computationally obtained wires is $A = 20$~\AA.
		}
		\label{fig:wire_melt_length}
	\end{figure}

	To evaluate the accuracy of our model, we now compare experimental measurements of nanowire conditioning to our computational model.
	Fig.~\ref{fig:wire_melt_length} shows the nanowire resistance as a function of applied conditioning current, for four different length wires.
	The experimentally measured resistance change is compared directly to simulated wires of the same length. 
	The material parameters used in the simulation are $D_{\rm{mean}} = 3.5$~\AA~and $D_{\rm{sd}} = 3.5$~\AA, which correspond well to the range shown in Fig.~\ref{fig:main_result} and the transmission electron microscopy. As the device to device variation is relatively large (see Fig.~\ref{fig:Histogram_resistance}), to compare to experiment we generate many simulated devices and choose one example with a total (un-conditioned) resistance similar to that seen in experiment.
	
	To model the experimentally observed IEM process, we use the parameter $A$ as a fitting parameter, finding that a value of $A=20$~\AA~provides good agreement for all four wires in Fig.~\ref{fig:wire_melt_length}. 
	Experimentally, the rate of change of the resistance with applied current depends on the length of the wire. We see that this behaviour is reproduced in our model, especially for the longer wire lengths.
	
\begin{figure}[t]
		\centering
		\includegraphics[width=\linewidth]{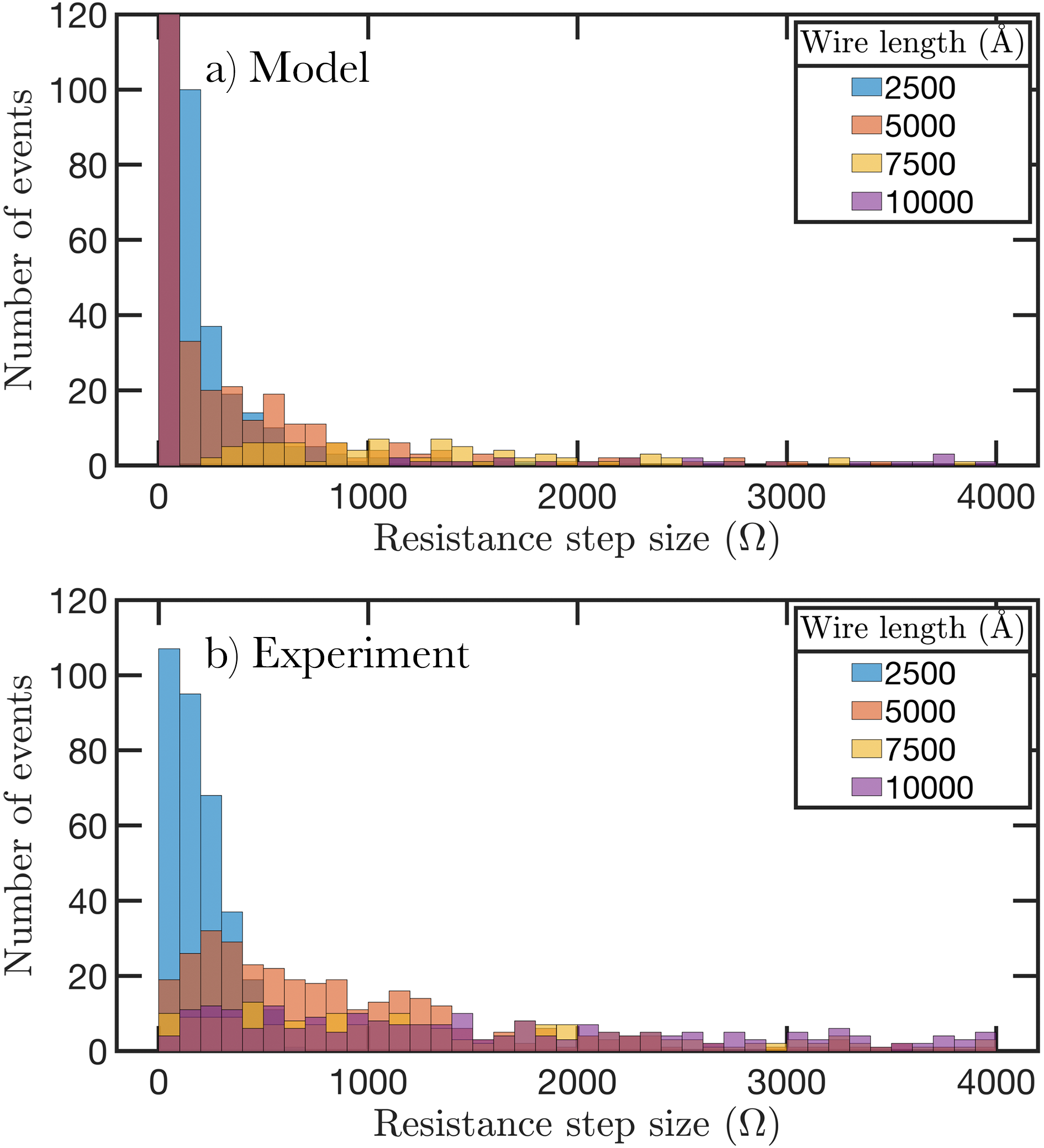}
		\caption{
		    Histogram of the resistance changes for the IEM processes shown in Fig.~\ref{fig:wire_melt_length}. 
			(a) Shows the resistance steps for the data obtained by the model outlined in this paper.
			(b) Shows the resistance steps as measured experimentally for the same length wires.
			Here we use a bin size of 100~$\Omega$.
		}
		\label{fig:Histogram_exp_and_model}
	\end{figure}

	Fig.~\ref{fig:Histogram_exp_and_model} shows histograms of the resistance steps during the IEM process. These show that both experimentally and in the computational model, the shorter wire lengths undergo smaller steps. This suggests that in the longer wires once the IEM process starts there are more opportunities for links to form, resulting in larger resistance steps. It is worth noting that there is a very large number of small ($<100$~$\Omega$) steps in the computational model (the vertical axis is truncated in Fig.~\ref{fig:Histogram_exp_and_model}a). Such small changes in resistance are not directly resolvable in the experiment where a variance of $>100$~$\Omega$ is observed even in the `flat' response at small applied currents.
	
	The agreement between theory and experiment is noteworthy given the simplicity of the model and strongly supports the idea that IEM is driven by the formation of conduction channels between grains. There are observed differences, particularly for shorter wires and higher currents, but this is not surprising given the idealised model of the contacts (as can be see in Fig.~\ref{fig:3_frame_melting}) which will have a non-negligible effect on the total resistance in the IEM calculations.	
	
	The idea that the IEM process is driven by QPC formation between grains provides us with hints on how to optimise the wire fabrication process. The dominant influence of the parameter $A$ in our model was to move the IEM curves in Fig.~\ref{fig:wire_melt_length} to lower applied currents as $A$ is increased. The linear relationship to stress duration $t$ implies increasing the duration of the current pulses will decrease the required current to achieve a given resistance reduction. Experimentally, we did not observe a dependence on the pulse length duration from 10ms up to minutes, which suggests that the QPC links form on a relatively short timescale (i.e. less than 10~ms) after which the microstructure is stable for a given drive current. As the parameter $A$ also depends on temperature, it may be possible to optimise the IEM process using current pulse magnitude, pulse duration and device temperature, however the details depend on the localised heating of the links and require more careful experimental study.

	\section{Conclusion} \label{section:conclusion}
	
	We have developed a network resistor model of the electronic properties of granular aluminium channels which is based on the material properties and microstructure of these films.
	Using this approach we can theoretically determine the sheet resistance and resulting nanowire resistance for a granular aluminium channel as a function of the device geometry and film characteristics.
	This model allows us to estimate the device to device variance and yield as a function of nanowire width and length, which is an important consideration when designing superconducting circuits based on such nanowires.
	Extending this model, we qualitatively explain the intrinsic electromigration within granular aluminium nanowires and suggest that it fundamentally results from current induced movement of oxygen atoms forming conduction channels between metallic grains within the film. Direct comparison to measurements of electromigration in nanowires of various lengths show good agreement for sensible choices of parameters.
	
	The uniformity if granular aluminium allows a high precision comparison between our computational model and the experimental measurements. However the details of the IEM process are more generally applicable in understanding the conduction properties of cermet films~\cite{deutscher2008nanostructured} and granular superconductors~\cite{portis1993electrodynamics,deutscher2006new}.
	Understanding the role microstructure and electromigration plays in the electrical response of superconducting nanowires provides a new approach to optimising fabrication and device design.
	
	\begin{acknowledgments}
		The authors acknowledge useful discussions with S. Brown. This research was supported by the Australian Research Council through grants CE170100039 and DP140100375. In addition, the work was supported by the Initiative and Networking Fund of the Helmholtz Association, the German BMBF Project PtQUBE and the Helmholtz International Research School for Teratronics (Y.S. and J.N.V.).
		The computational modelling was undertaken with the assistance of resources from the National Computational Infrastructure, which is supported by the Australian Government.
	\end{acknowledgments}
		
	\bibliography{granal_references.bib}
	
\end{document}